\documentclass[useAMS,usenatbib,usegraphicx]{mn2e}
\usepackage{soul}
\usepackage{graphicx}
\usepackage{color}
\usepackage{float}

\def\axpu{AXP~4U~0142+61~}
\def\rxs{1RXS~J170849.0-400910~}

\def\kes{1E~1841-045~}
\def\kesnos{1E~1841-045}
\def\axpe{1E~2259+586}
\def\axpes{1E~2259+586~}
\def\1e1048{1E~1048.1-5937}
\def\CXOU{CXOU~J164710.2-455216~}

\title{A glitch and an anti--glitch in the anomalous X-ray pulsar \kes} 

\author[S. \c{S}a\c{s}maz Mu\c{s} et al.]{S. \c{S}a\c{s}maz Mu\c{s}$^{1}$\thanks{E-mail:
sinemsm@sabanciuniv.edu}, Berk Ayd{\i}n$^{1}$, and Ersin G\"o\u{g}\"u\c{s}$^{1}$\\
$^{1}$Sabanc\i\ University, Orhanl\i - Tuzla, \.Istanbul, 34956, Turkey}
\begin{document} 

\pagerange{\pageref{firstpage}--\pageref{lastpage}} \pubyear{2014}
\maketitle

\label{firstpage}


\begin{abstract}  

We investigated the long-term spin properties of the 
anomalous X-ray pulsar (AXP) \kes by performing a 
temporal analysis of archival {\it RXTE} observations 
spanning about 5.2 yr from 2006 September to 2011 December. 
We identified two peculiar timing anomalies within 
$\sim$1 yr of each other: a glitch with 
$\Delta{\nu}$/$\nu$$\sim$4.8 $\times$ 10$^{-6}$ near 
MJD 54303; and an anti-glitch with 
$\Delta{\nu}$/$\nu$$\sim$ $-$5.8 $\times$ 10$^{-7}$ near 
MJD 54656. The glitch that we identified, which is the
fourth glitch seen in this source in the 13 yr of RXTE 
monitoring, is similar to the last two detected glitches. 
The anti-glitch from \kesnos, however, is the first to be 
identified. The amplitude of the anti-glitch was comparable 
with that recently observed in AXP \axpe. We found no 
significant variations in the pulsed X-ray output of the 
source during either the glitch or the anti-glitch. We 
discuss our results in relation to the standard pulsar 
glitch mechanisms for the glitch, and to plausible 
magnetospheric scenarios for the anti-glitch.

\end{abstract} 
 

\section{Introduction} 

Anomalous X-ray pulsars (AXPs), along with soft 
gamma-ray repeaters (SGRs) are extremely magnetized 
neutron stars (magnetars) powered by the decay of
their strong magnetic fields. These sources exhibit 
numerous unusual characteristics, such as, relatively 
slow rotation speeds, high spin-down rates, bright 
persistent X-ray emission and, for most of them, 
episodic bursts seen in X-rays \citep[see e.g.,]
[for recent reviews]{reaesposito11,mereghetti11}. 
The dipole magnetic field strengths, inferred 
from their spin periods and spin down rates are 
indeed extremely high, which are sufficient to 
account for their observed unique properties 
\citep{dunthomp92}. 

Long term spin behaviour of magnetars usually do 
not follow a secular trend, likely due to large 
magnetic torques along with episodic wind outflow 
that could take place in strongly magnetized 
environments \citep{thompson00}. Additionally, 
sudden increase in the angular velocity 
(i.e., glitches) has been observed from several 
AXPs. However, glitch events in AXPs have peculiar 
differences compared to the properties of glitches 
from rotation powered pulsars: \1e1048 is one of 
the most variable AXP both in timing and radiative 
behaviour as several short energetic burst and 
flare events were observed 
\citep{gavriil02,tam08,dib09,gavriil06}. Its 2007 
flare event was coincident with a large glitch 
\citep{dib09}. Similarly, \axpu went into an active 
period, exhibiting six bursts and a glitch event 
which was over-recovered, causing the neutron star 
to rotate slower than the pre-glitch level 
\citep{gavriil11}. Another AXP showing coincident 
burst and glitch event is \CXOU 
\citep{krimm06,israel07b,muno07,woods11}. On the 
other hand, \rxs and \kes have experienced glitches, 
but there has been no evidence of accompanying 
radiative enhancements in these sources \citep{kaspi00,kaspi03b,dosso03,israel07a,dib08,ssmeg13}.

In \axpe, an outburst has been associated with a 
glitch \citep{kaspi03a,woods04}. More interestingly, 
this source exhibited a sudden decrease in its 
angular velocity, namely an anti-glitch 
\citep{archibald13a}, within two weeks of a hard
X-ray burst \citep{foley12}. The sudden spin-down 
trend in this AXP occurred in conjunction with a 
doubled persistent source flux, as reported by 
\citep{archibald13a}. These authors also report 
that the observed anti-glitch was followed by 
either a glitch event $\sim$90 d later or by a 
second anti-glitch $\sim$50 d apart which was
suggested as a more plausible explanation based on 
a Bayesian approach \cite{hu13}. Unlike glitches 
observed from isolated neutron stars, which are 
generally attributed to an internal mechanism 
\citep[e.g,][]{anderson75,alpar77}, the anti-glitch
event was explained in terms of external effects, 
including magnetospheric processes 
\citep{lyutikov13,tong14,katz14} or the accretion 
of orbiting objects \citep{katz14,ouyed13,huang14}.

\kes has a pulse period of $\sim$11.8 s. It is a 
bright persistent X-ray source with an emission 
spectrum extending into hard X-rays, up to about 
150 keV \citep{kuiper04}. It is also a source of 
several short energetic magnetar bursts 
\citep{kumarsafi10,lin11,collazzi13,palshin13}. 
Three glitches have been observed from \kes with 
amplitudes ranging from $\sim$10$^{-7}$ Hz to 
1.2$\times$10$^{-6}$ Hz \citep{dib08}. Note that 
these events were radiatively silent, i.e, there 
was no significant variations of the radiative 
behaviour of the source associated with these 
timing anomalies \citep{dib08,zhu10}. The 
persistent X-ray emission of \kes has remained 
constant during the duration of energetic bursts 
\citep{lin11,archibald13b}.

Here, we present the results of long-term timing 
analysis of \kes using Rossi X-ray Timing Explorer 
({\it RXTE}) observations spanning $\sim$5.5 years. 
In the following section, we introduce these 
{\it RXTE} observations and our data analysis 
scheme. In \S\ref{sect:results}, we present the 
results of our detailed temporal investigations, 
and report on the discovery of an anti-glitch and 
additional glitch from this source. We then 
discuss our findings in \S\ref{sect:discuss}.


\begin{table*}
\centering
\begin{minipage}{175mm}
\caption{Pulse Ephemeris of \kesnos$^{a}$}
\begin{tabular}{lcccccc}
\hline
Parameters Name&Segment 0&Segment 1$^{b}$ & Segment 2$^{b}$ &Segment 3&Segment 4&Segment 5 \\
\hline
Range (MJD)  & 53829 $-$ 54076 & 54126 $-$ 54431 & 54492 $-$ 54807 & 54860 $-$ 55168 & 55223 $-$ 55538 & 55588$-$ 55903 \\
Epoch (MJD) & 53823.9694 & 54125.967 &54491.992 & 54860.107 & 55223.090 & 55587.871 \\
Number of TOAs   & 19 & 26 & 24 & 22 & 29 & 24 \\
$\nu$ (Hz)  & 0.084868766(5) & {\it 0.084861458(8)} &{\it 0.084852295(8)} & 0.084843233(2)& 0.084834325(4) & 0.084824920(3) \\
$\dot{\nu}$  ($10^{-13}$ Hz s$^{-1}$)  & $-$2.82(1) &{\it $-$3.92(2)} &{\it $-$2.72(2)} & $-$2.834(1) & $-$2.883(7)& $-$2.943(6) \\
$\ddot{\nu}$ ($10^{-22}$ Hz s$^{-2}$)  & $-$5.6(9) &{\it 187(4)} &{\it $-$23(4)} & $-$ & $-$9.0(5) & $-$4.1(4)\\
$d^{3}\nu/dt^{3}$ ($10^{-28}$ Hz s$^{-3}$)  & $-$ &{\it $-$12.4(3)} &{\it 1.5(3)} & $-$ & $-$ & $-$\\
rms (phase) & 0.0143 &{\it 0.1184} &{\it 0.0210} & 0.0160 & 0.0230 & 0.0226 \\
$\chi$${^2}$/DOF & 16/15 &{\it 1508.6/21} &{\it 37.8/19} & 19.1/19 & 46.6/25 & 40.6/20 \\
\hline
\end{tabular}
\medskip
\\
$^{a}$ Values in parentheses are the uncertainties in the last digits of their associated measurements.\\
$^{b}$ These spin parameters yield unacceptable fits to data but listed here to illustrate the inadequacy of the polynomial model.
\end{minipage}
\label{tab:tablemain}
\end{table*} 


\section{Observations and Data Processing}
\label{sect:analysis}

\kes has been observed with {\it RXTE} in 279 
occasions over a time span of $\sim$13 years from 
1999 February to 2011 December\footnotemark 
\footnotetext{The complete list of {\it RXTE} 
observations as well as pointing details can be 
obtained from the High Energy Astrophysics Science 
Archive Research Center of NASA at 
http://heasarc.gsfc.nasa.gov.}. Data covering the 
first $\sim$7.6 years have already been investigated 
by \cite{dib08}. Here, we investigated 137 
{\it RXTE} observations performed from 2006 September 
19 to 2011 December 8 for the first time. Additionally, 
we also included the last 12 (2006 April 3 - 2006 
September 5) {\it RXTE} pointings in the sample of 
\cite{dib08} in order to link our long term timing 
results with their extensive coverage. Exposure times 
of these 149 pointings were between $\sim$1.2 ks and 
$\sim$9.6 ks with a mean of $\sim$ 4.6 ks and spacing 
between successive pointings varied between 0.04 and 
61 days with an average of 14 days. 

We employed data collected with the Proportional 
Counter Array (PCA), that was an array of nearly 
identical five proportional counter units (PCUs) 
operated optimally in the energy range of 2$-$30 
keV \citep{jahoda06}. We first filtered each 
observation for occasional bursts, data anomalies 
and instrumental rate spikes by screening their 
light curves in the 2-30 keV band with the 31.25 
ms time resolution. We then converted the arrival 
times of the remaining events to the Solar system
barycenter. In order to maximize the signal-to-noise 
ratio for timing analysis of \kesnos, we selected 
events in the 2$-$11 keV energy range recorded  
at the top Xenon layer of each operating PCU in 
GoodXenon mode, as was also done by \cite{dib08}.


\section{Data Analysis \& Results}
\label{sect:results}

In order to undertake coherent timing analysis, we 
grouped all 149 observations into six segments 
intercepted with observational interruptions in 
between due to Solar constraints. In addition, we 
merged observations together if the time spacing 
between them was less than 0.1 d. In this grouping 
scheme, segment 0 has 12 observations which are 
the last {\it RXTE} pointings used by \cite{dib08}. 
We first generated a high signal-to-noise ratio 
pulse profile template using a subset of observations 
(typically 5-6), that were performed in the beginning 
of each segment. We obtained the pulse profile for 
each observation by folding light curves with a 
nominal spin frequency. We then cross-correlated the 
pulse profiles with the template and measured their 
phase shifts with respect to the template. Finally,
we fitted the phase shifts with a polynomial of 
the following form:

\begin{equation}
\phi(t) = \phi_{0}(t_{0})+\nu_{0}(t-t_{0})+ \frac{1}{2}\dot{\nu_{0}}(t-t_{0})^{2} + ...
\label{eq:taylorphase}
\end{equation}
where t$_{0}$ is the epoch time.
We found that all segments, except for Segment 1 
and Segment 2, are fitted well with polynomials 
of the third order or lower. We present the 
results of polynomial model fits to each segment 
in Table~\ref{tab:tablemain}. We note that 
the spin frequency and spin-down rate of Segment 
0 are consistent with those reported by \cite{dib08}. 
In Table~\ref{tab:tablemain}, we also list 
root-mean-square (rms) fluctuations of the resulting 
phase residuals, which are presented in the top 
panel of Figure~\ref{fig:resfreqpcount}. A 
fourth-order polynomial fit to pulse arrival times 
in Segment 1 yields extremely large fit statistics 
($\chi^{2}$  of 1508.6 for a degrees of freedom 
(DOF) of 21; see Figure~\ref{fig:resfreqpcount} top 
panel and Figure~2 panel b). We therefore identified 
this segment to search for glitch(es), as we describe 
in detail below.

We fitted the pulse arrival times of Segment 2 with 
a fourth-order polynomial but obtain unacceptable 
fit statistics ($\chi^{2}$/DOF = 37.8/19). We note 
the fact that there are systematic variations of 
the phase residuals around MJD 54650. Higher 
order polynomial fit to this segment results in 
lower fit statistics, whereas the errors of derived 
spin parameters become large. For these reasons, we 
also investigated this segment for searching timing 
anomalies.

We found that arrival times of Segment 3 can be 
represented with a second-order polynomial, that is 
the lowest order we obtained among our investigation 
span of 5.5 yr. We also find that the frequency 
derivatives\footnotemark\footnotetext{Frequency 
derivatives are obtained by fitting a second order 
polynomial to subset of observations of $\sim$2 
months long} within this segment remain constant 
(see the middle panel in Figure~\ref{fig:resfreqpcount}).   
For Segments 4 and 5, we obtained an adequate fit 
with a third-order polynomial. However, the rms phase 
residual fluctuations for these segments are larger 
than other segments, namely Segments 0 and 3. Note 
the important fact that energetic short duration 
bursts from this magnetar were observed during 
Segments 4 and 5, as they are denoted with vertical 
solid lines in Figure~\ref{fig:resfreqpcount}.


\setcounter{figure}{0}
\begin{figure*}
\vspace{0.0in}
\begin{center}
\includegraphics[scale=0.8]{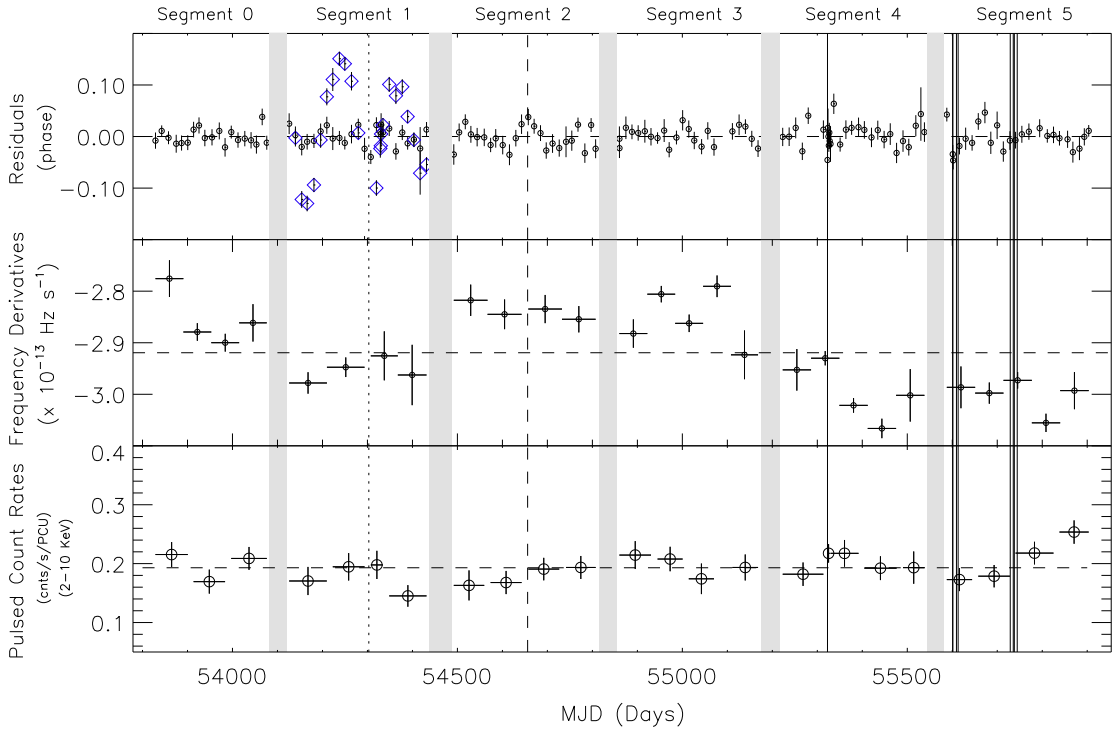}
\vspace{0.9cm}
\caption{Top panel: phase residuals after the subtraction of the polynomial 
models presented in Table~\ref{tab:tablemain} from data. 
For Segment 1 we present the phase residuals after subtraction of the 
polynomial model (blue diamonds) presented in Table~\ref{tab:tablemain} and 
glitch model presented in Table~\ref{tab:glitch4} (See Figure~2 for
more detailed presentation of this segment).  
Middle panel: frequency derivative evolution of the source.
Bottom panel: evolution of the pulsed count rates in the 2$-$10 keV band. 
The times of glitch 4 and anti-glitch are shown with the vertical doted and 
dashed lines, respectively. The solid vertical lines indicate the times of energetic
bursts listed in table 1 of \protect\cite{lin11}. Data gaps are indicated with
light grey bars.}
\label{fig:resfreqpcount}
\end{center}
\end{figure*}

In order to determine whether the large fluctuations 
in the phase residuals of Segments 1, 2, 4 and 5 are 
due to a sudden change in the spin frequency of the 
source, we fitted phase shifts of these segments 
using MPFITFUN routine \citep{markwardt09} with a 
model involving a quadratic polynomial and a sudden 
change of spin frequency (i.e., glitch). 
The corresponding spin trend of this model is:
\begin{equation}
\nu(t) = \nu_{0}(t) + \Delta{\nu} + \Delta{\dot{\nu}}(t - t_{g})
\label{eq:glt}
\end{equation}
Here $t_{g}$ is the time of the glitch. $\Delta{\nu}$ 
is the change in the frequency at the time of the glitch. 
$\Delta{\dot{\nu}}$ is the frequency derivative change 
after the glitch and $\nu_{0}(t)$ is pre-glitch frequency
evolution.

In Segment 1, we found that a model involving a glitch 
provides statistically significant improvement in fitting 
the phase shifts ($\chi^{2}$/DOF = 31.6/21). We, therefore, 
conclude that there is a glitch at MJD $\sim$54303 with 
an amplitude of $\Delta\nu$ $\sim$4$\times$10$^{-7}$ Hz. 
This is the 4th glitch observed from \kesnos. 
We present the parameters for Glitch 4 in 
Table~\ref{tab:glitch4}. Note that the amplitude of this 
glitch is on the order of last two glitches observed 
from this source \citep[see][]{dib08}. We also present 
the phase residuals of the polynomial model and glitch 
model for comparison in Figure~2.


\begin{figure}
\vspace{0.0in}
\hspace{0.5cm}
\includegraphics[width=8.5cm]{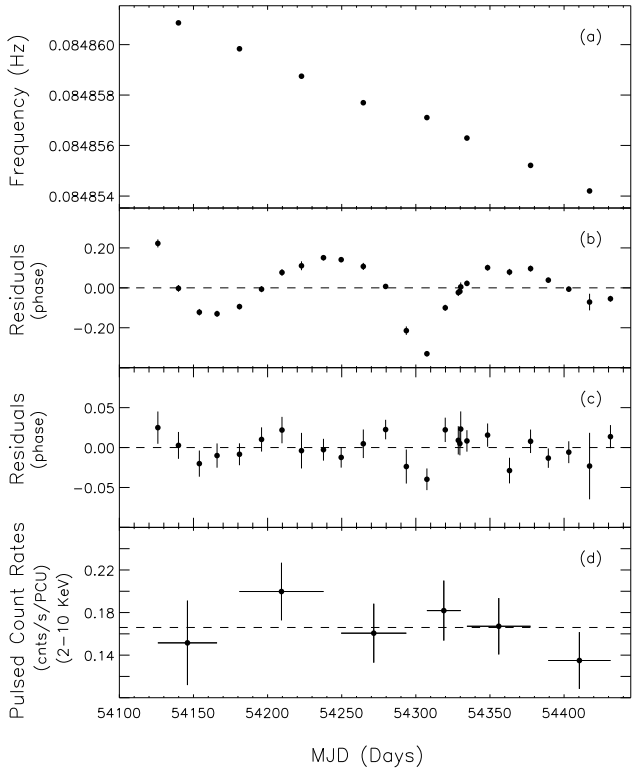}
\vspace{-4.0cm}
\caption{(a) Spin frequency evolution of the source during Segment 1. 
(b) Phase residuals after the subtraction of a fourth order polynomial model
from the data. (c) Phase residuals after subtracting the glitch model. 
(d) Pulsed count rates of the source in the 2$-$10 keV band obtained using 
sets of observations spanning mostly over 40$-$50 d to ensure significant
pulsed flux measurement.} 
\label{fig:glitch4}
\end{figure}


\begin{table}
\centering
\begin{minipage}{175mm}
\caption{Parameters for Glitch 4$^{a}$}
\label{tab:glitch4}
\begin{tabular}{lc}
\hline
Range (MJD)&54126$-$54431\\
Epoch (MJD)&54125.967 \\
Number of TOAs&26\\
$\nu$ (Hz)&0.084861236(2)\\
$\dot{\nu}$ ($10^{-13}$ Hz s$^{-1}$)&$-$2.965(3)\\
$t_{g}$ (MJD)&54303(3)\\
$\Delta{\nu}$ ($10^{-8}$ Hz)&40.7(6)\\
$\Delta{\dot{\nu}}$ ($10^{-15}$ Hz s$^{-1}$)&1.2(7)\\
rms (phase)&0.0174 \\
$\chi$${^2}$/DOF &31.6/21 \\
\hline
\end{tabular}
\medskip
\\
$^{a}$ Values in parentheses are the uncertainties \\
in the last digits of their associated measurements.
\end{minipage}
\end{table}

 
\begin{table}
\begin{minipage}{175mm}
\caption{Parameters for Anti-glitch$^{a}$}
\begin{tabular}{lc}
\hline 
Range (MJD)&54492$-$54807\\
Epoch (MJD)&54491.992 \\
Number of TOAs&24\\
$\nu$ (Hz)&0.084852317(2)\\
$\dot{\nu}$ ($10^{-13}$ Hz s$^{-1}$)&$-$2.833(3)\\
$t_{g}$ (MJD)&54656.0\\
$\Delta{\nu}$ ($10^{-8}$ Hz)&$-$4.9(6)\\
$\Delta{\dot{\nu}}$ ($10^{-15}$ Hz s$^{-1}$)&$-$1.8(6)\\
rms (phase)&0.0139 \\
$\chi$${^2}$/DOF &16.3/19\\
\hline
\end{tabular}
\medskip
\\
$^{a}$ Values in parentheses are the uncertainties \\
in the last digits of their associated measurements.
\end{minipage}
\label{tab:antiglitch}
\end{table} 

\begin{figure}
\vspace{0.0in}
\hspace{0.5cm}
\includegraphics[width=8.5cm]{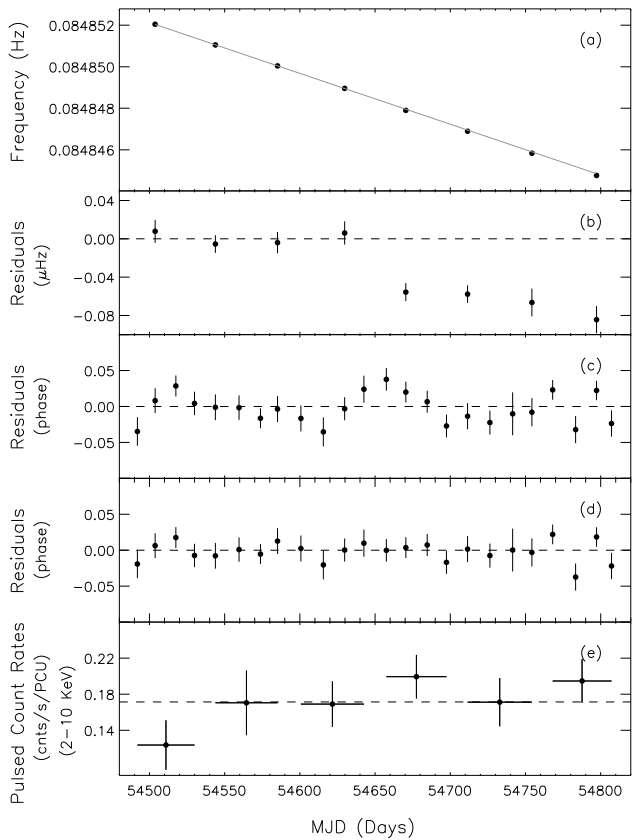}
\label{fig:antiglitch}
\vspace{-2.0cm}
\caption{(a) Spin frequency evolution of the source during Segment 2. The solid 
line is the spin-down trend obtained by fitting the observations before 
MJD $\sim$54630, and extrapolated onwards. 
(b) Residuals of the model and its extrapolation in (a). 
(c) Phase residuals after the subtraction of a fourth order 
polynomial fit presented in Table~\ref{tab:tablemain}.
(d) Phase residuals after the subtraction the glitch model presented 
in Table~\ref{tab:antiglitch}
(e) The pulsed count rates in 2$-$10 keV averaged over $\sim$40 d.}
\end{figure}

As we noted earlier, pulse phase modeling of the 
Segment 2 with a fourth order polynomial yields an 
unacceptable fit statistics (see panel c of 
Figure~\ref{fig:antiglitch}), while increasing the 
order of polynomial results in unconstrained  spin 
parameters. For this reason, we also modeled the 
phases of this segment with a glitch model. We find 
that an ordinary glitch model fit does not improve 
the fit statistics. However, a glitch with a 
negative amplitude (i.e., an anti-glitch) of 
$\Delta\nu$ $\sim$$-$5$\times$10$^{-8}$ Hz provides 
a significant improvement in fit statistics 
($\chi^{2}$/DOF = 16.3/19); a $\Delta\chi^2$ of 21.5 
for the same number of DOF as the polynomial model 
fit. Based on this, we conclude that \kes exhibited 
an anti-glitch near MJD 54656, that is $\sim$1 yr 
after Glitch 4. We list the parameters of the 
anti-glitch model fit in Table~\ref{tab:antiglitch}.
Note that fit results suggesting the anti-glitch 
epoch between MJD 54645 and 54662 yield similar fit
statistics, indicating that the anti-glitch occurred
in this time interval. In the top panel of 
Figure~\ref{fig:antiglitch} we present the frequency 
evolution of the source in this segment. To show the 
sudden deviation of the spin frequency, we fit a 
linear trend to the frequencies prior to MJD 
$\sim$54650 and extrapolated this fit to the rest of 
this segment (see panel b in Figure~\ref{fig:antiglitch}). 
We found that the spin-down rate before MJD 54630 and 
after MJD 54670 are consistent with one another
($-$2.84(1)$\times$10$^{-13}$ and 
$-$2.86(1)$\times$10$^{-13}$ Hz~s$^{-1}$, respectively). 
The average spin-down rate during the $\sim$40 d in 
between is about $-$3$\times$10$^{-13}$ Hz~s$^{-1}$. 
We also present the phase residuals of the anti-glitch 
involving model in the panel d of Figure~
\ref{fig:antiglitch}.


Fits to the pulse arrival times of Segment 4 and 
5 with models involving a glitch or anti-glitch 
did not yield any improvement in the fit statistics. 
We therefore conclude that the system exhibits 
higher level of timing noise, likely related to 
the emission of numerous energetic bursts
during these episodes (see Figure 
\ref{fig:resfreqpcount}).

Finally, in order to search for radiative variabilities 
we performed pulsed count rate analysis as explained 
in \cite{ssmeg13}. We found that pulsed count rate of 
the source is constant over $\sim$5.5 yr of
{\it RXTE} observations (See, third panel of Figure~
\ref{fig:resfreqpcount}).


\section{Discussion and Conclusions}
\label{sect:discuss}

\kes was observed $\sim$13 yr by {\it RXTE}. Here 
we performed timing analysis of \kes using $\sim$5 
yr of {\it RXTE} observations. Previous $\sim$7.6 yr 
has been analysed by \citep{dib08} and 3 glitches 
have been reported. The largest glitch observed from 
this source has an amplitude of $\sim$1.2$\times$10$^{-6}$ 
Hz with a recovery timescale of 43 days and fractional 
increase of 0.1 in its spin-down rate. Consequent two 
glitches have amplitudes on the order of $\sim$10$^{-7}$ 
Hz without any observable exponential recovery. There is 
no evidence of accompanying radiative enhancements 
during all three glitch epochs.

Through our detailed investigations of {\it RXTE} 
monitoring of \kes spanning over five years, we have 
identified two timing events separated by $\sim$1 yr: 
a glitch and an 'anti-glitch'. The glitch event has 
occurred at MJD $\sim$54303 with an amplitude of 
$\Delta\nu$$\sim$4$\times$10$^{-7}$ Hz without any 
observable exponential recovery. Note that this is the 
fourth glitch identified from this source: one of the 
earlier three has an amplitude of $\sim$10$^{-6}$ Hz 
with a recovery timescale of 43 days \citep{dib08}, 
while the other two were at similar amplitudes and 
showed no exponential recovery. Similar to the three 
earlier glitch episodes in \kes, we found no associated 
radiative enhancement deduced from the pulsed X-ray 
flux measurement (bottom panels of Figure~
\ref{fig:resfreqpcount} and 2). 

The second event, an anti-glitch seen from \kes for 
the first time, has occurred at MJD $\sim$54656 with 
an amplitude of $\Delta\nu$$\sim$$-$5$\times$10$^{-8}$ 
Hz. The amplitude of the only other anti-glitch event 
observed from \axpes \citep{archibald13a} is strikingly 
similar. We found that the source experienced an 
elevated spin-down rate over about 40 days, and 
$\dot{\nu}$ returned back to the pre-anti-glitch level 
after MJD 54670. Note that the average spin-down 
rate in the elevated regime is about $-$3$\times$$10^{-13}$ 
Hz~s$^{-1}$, which is similar to the $\dot{\nu}$ values 
exhibited by the source over $\sim$150 d following an 
energetic burst on MJD 55322.6 (see the middle panel of 
Figure~\ref{fig:resfreqpcount}). Both pulsed X-ray 
emission (see the bottom panel of Figure~\ref{fig:antiglitch}) 
and the 0.5-10 keV flux \citep[see figure 3 of][]{lin11} 
of the source remain constant during the anti-glitch episode, 
similar to the case for the fourth glitch.

According to the standard pulsar glitch models 
\citep[see for example][]{anderson75,alpar77}, a faster 
rotating superfluid transfers angular momentum to the 
crust which results in positive increment in the 
observed spin frequency of the neutron star. \cite{alpar94} 
statistically determined that superfluid vortex unpinning 
model with a constant fractional vortex density, which 
is the fraction of vortex density involved in the glitch 
event, provides the most plausible explanation for glitches 
observed from rotation powered pulsars. In this model, 
the number of glitches in a given time span that a pulsar 
would experience is related to the constant fractional 
vortex density, the ratio of the spin-down rate and frequency 
of the pulsar. From previous observations and glitches 
observed from \kes we determined the fractional vortex density 
as 2.90$\times$10$^{-4}$. Note that this is consistent with 
fractional vortex density for \rxs \citep{ssmeg13}. Using this 
parameter for \kes and an average value of the ratio of the 
spin-down rate to the frequency yield the expected number of 
glitches from this magnetar during the entire 13 yr of 
{\it RXTE} observation span as 4.5. With the fourth glitch 
we uncovered in this study, our results are in agreement 
with the expectations of the vortex unpinning model.

Recent observation of the anti-glitch from \axpes 
implies a neutron star superfluid interior rotating slower 
than the crust within the context of standard pulsar 
glitch models \citep{archibald13a, anderson75}. It was 
suggested that the rotation of the superfluid can be 
slowed down by crustal deformations due to magnetic 
stresses in highly magnetized sources \cite{thompson00,duncan13}.
\cite{thompson00} suggest that \axpes and \kes might 
have had episodes of  accelerated spin-down due to the 
fact that they are older than their associated supernovae 
remnants as inferred by their characteristic ages.
In \cite{thompson00} they also propose a particle outflow 
scheme to account for the sudden spin-down behaviour of 
SGR 1900+14\footnotemark\footnotetext{The rapid spin-down 
trend in SGR 1900+14 can be considered as the first 
observational manifestation of an anti-glitch, though 
the sparsity of observations did not allow the firm 
confirmation.} \citep{woods99}, in conjunction with 
its August 27 giant flare. However, there was no 
indication of particle outflow from \axpes around the 
time of the anti-glitch \citep{archibald13a}. For this 
reason, the particle outflow scenario was discarded. 
Alternatively, there are already a couple of 
magnetospheric models suggested to understand the 
origin of the anti-glitch: \cite{lyutikov13} proposed that 
the sudden spin-down and in general variable spin-down trends 
are caused by the changes in torque due to transient opening 
of a small region of the twisted magnetosphere during the 
X-ray burst. \cite{tong13,tong14} have applied the wind 
braking scenario \citep{michel69,harding99,thompson00} to 
magnetars in the case that the rotational energy of the star 
is mainly extracted via a constant particle wind from the star. 
In this model an anti-glitch corresponds to an enhanced state 
of the particle wind \citep{tong14}. Both partial 
magnetospheric opening and wind breaking models require 
radiative enhancements accompanying the anti-glitch, 
which was the case for \axpes as its X-ray flux increased 
by at least a factor of two in association with the 
anti-glitch \cite{archibald13a}. Our results, however, 
place an indirect constraint on both models since we found 
no observable variations in the pulsed X-ray emission from 
\kes at the time of its anti-glitch. Nevertheless, occasional 
detection of energetic bursts from \kes indicate that its 
magnetosphere is active and not all but some X-ray bursts 
may lead to a magnetospheric rearrangement that could lead 
to the episodic rapid spin-down as prescribed by 
\cite{lyutikov13}.

Note the fact that we became aware of the paper by 
\cite{dib14} during the review stage of our paper. They 
fit a very wide data segment (spanning more than 1200 d 
and including our suggested anti-glitch) with a fifth-order 
polynomial, and find an rms of 0.041, which is much 
larger than that for any other segment. Even with the 
fifth-order polynomial fit, large fluctuations in the 
phase residuals are clearly visible (see panel (c) around 
MJD 54600 of figure 1c in \cite{dib14}). They make no 
emphasis on this significant timing anomaly, which we 
interpret as an anti-glitch.

\section*{Acknowledgements}
We thank M. Ali Alpar for helpful comments.
S\c{S}M acknowledges support through the national 
graduate fellowship program of The Scientific and 
Technological Research Council of Turkey 
(T\"UB\.{I}TAK).

\end{document}